\documentclass[prl,twocolumn,superscriptaddress,amsfonts,amssymb,showpacs,a4]{revtex4}
\usepackage{graphicx}% Include figure files
\usepackage{dcolumn}% Align table columns on decimal point
\usepackage{times}
\usepackage{amsmath}
\usepackage{upgreek}

\usepackage{color}

\begin{document}
\title{Strong ferromagnetism at the surface of an antiferromagnet caused by buried magnetic moments}

\author{A. Chikina}
\affiliation{Institute of Solid State Physics, Dresden University of Technology, D-01062 Dresden, Germany}

\author{M. H\"oppner}
\affiliation{Max Planck Institute for Solid State Research, Heisenbergstrasse 1, D-70569 Stuttgart, Germany}
\affiliation{Institute of Solid State Physics, Dresden University of Technology, D-01062 Dresden, Germany}

\author{S. Seiro}
\affiliation{Max Planck Institute for Chemical Physics of Solids, D-01187 Dresden, Germany}

\author{K. Kummer}
\affiliation{European Synchrotron Radiation Facility, 6 Rue Jules Horowitz, Bo\^{\i}te Postale 220, F-38043 Grenoble Cedex, France}

\author{S. Danzenb\"acher}
\affiliation{Institute of Solid State Physics, Dresden University of Technology, D-01062 Dresden, Germany} 

\author{S. Patil}
\affiliation{Institute of Solid State Physics, Dresden University of Technology, D-01062 Dresden, Germany} 

\author{M. G\"uttler}
\affiliation{Institute of Solid State Physics, Dresden University of Technology, D-01062 Dresden, Germany}

\author{Yu. Kucherenko}
\affiliation{Institute of Solid State Physics, Dresden University of Technology, D-01062 Dresden, Germany}
\affiliation{Institute for Metal Physics, National Academy of Sciences of Ukraine, UA-03142 Kiev, Ukraine}

\author{E. V. Chulkov}
\affiliation{Departamento de Física de Materiales, Facultad de Ciencias Químicas, Universidad del País Vasco, Apartado 1072, 20080 San Sebastián/Donostia, Spain}
\affiliation{Tomsk State University, 634050 Tomsk, Russia}

\author{Yu. M. Koroteev}
\affiliation{Tomsk State University, 634050 Tomsk, Russia}

\author{K. K\"opernik}
\affiliation{Leibniz Institute for Solid State and Materials Research, IFW Dresden, D-01069 Dresden, Germany}

\author{C. Geibel}
\affiliation{Max Planck Institute for Chemical Physics of Solids, D-01187 Dresden, Germany}

\author{M. Shi}
\affiliation{Swiss Light Source, Paul Scherrer Institute, CH-5232 Villigen-PSI, Switzerland}

\author{M. Radovic}
\affiliation{Swiss Light Source, Paul Scherrer Institute, CH-5232 Villigen-PSI, Switzerland}
\affiliation{Laboratory for Synchrotron and Neutron Spectroscopy-ICMP, Ecole Polytechnique Federale de Lausanne, CH-1015 Lausanne, Switzerland}

\author{C. Laubschat}
\affiliation{Institute of Solid State Physics, Dresden University of Technology, D-01062 Dresden, Germany}

\author{D. V. Vyalikh}
\affiliation{Institute of Solid State Physics, Dresden University of Technology, D-01062 Dresden, Germany}

\date{\today}
\begin{abstract}

Carrying a large, pure spin magnetic moment of 7 $\upmu$B/atom in the half-filled 4f shell, divalent europium is an outstanding element for assembling novel magnetic devices in which a two-dimensional electron gas (2DEG) is polarized due to exchange interaction with an underlying magnetically-active Eu layer, even in the absence of a magnetic field.   
A natural example for such geometry is the intermetallic layered material EuRh$_2$Si$_2$, in which magnetically active layers of Eu are well separated from each other by non-magnetic Si-Rh-Si trilayers. Applying angle-resolved photoelectron spectroscopy (ARPES) to this system, we discovered a large spin splitting of a Shockley-type surface state formed by itinerant electrons of the Si-Rh-Si surface related trilayer. ARPES shows that the splitting sets in below approx. 32.5~K and quickly saturates to around 150~meV upon cooling. Interestingly, this temperature is substantially higher than the order temperature of the Eu 4f moments ($\approx$ 24.5~K) in the bulk. 
Our results clearly show that the magnetic exchange interaction between the surface state and the buried 4f moments in the 4$^{\mathrm{th}}$ subsurface layer is the driving force for the formation of itinerant ferromagnetism at the surface. We demonstrate that the observed spin splitting of the surface state, reflecting properties of 2DEG, is easily controlled by temperature. Such a splitting may also be induced into states of functional surface layers deposited onto the surface of EuRh$_2$Si$_2$ or similarly ordered magnetic materials with metallic or semiconducting properties. 

\end{abstract}


\maketitle
For a long time, rare-earth (RE) intermetallic materials have attracted considerable interest because of their exotic properties at low temperatures which include complex magnetic phases, valence fluctuations, heavy-fermion states, Kondo behavior and many others~\cite{1,2,3,4,5,6}. Europium with its half-filled 4f~shell has a unique position among the lanthanides. For the free Eu~atom and Eu~metal, the 4f$^7$ configuration -- corresponding to a divalent Eu~state -- is the stable one, but in many intermetallic compounds the Eu~4f shell is occupied by six electrons only. Since both configurations are nearly degenerate in energy, tiny changes in stoichiometry, doping or external parameters like pressure and temperature may lead to transitions from a divalent \mbox{[Xe]4f$^7$(5d6s)$^2$} to a trivalent \mbox{[Xe]4f$^6$(5d6s)$^3$} state of Eu, or even stabilize mixed-valent behavior. According to Hund’s rules the 4f$^6$ configuration is a Van-Vleck paramagnet with zero total effective angular momentum in the ground state, while the divalent 4f$^7$ configuration reveals a large pure spin momentum ($J = S = 7/2$) and can give rise to sophisticated magnetic properties in Eu-based intermetallics.
In this regard, EuRh$_2$Si$_2$ is apparently well-suited to give insight into the interplay between magnetic and electronic degrees of freedom~\cite{7}. Here, Eu exhibits a stable 4f$^7$ configuration (Eu$^{2+}$) from low temperature to room temperature with a pure spin magnetic moment of 7~$\upmu$B. It crystallizes in the tetragonal body-centered ThCr$_2$Si$_2$ structure~\cite{7, 8}, similarly to the heavy-fermion YbRh$_2$Si$_2$ system~\cite{9, 10} or the famous ”hidden order” material URu$_2$Si$_2$~\cite{11}. Its electronic structure is strongly correlated, involving massless and heavy quasiparticles which are mutually interacting~\cite{12}. Below the Néel temperature $T_{\mathrm{N}} = 24.5$~K the Eu~4f moments in EuRh$_2$Si$_2$ order antiferromagnetically (AFM)~\cite{7}. Like in several other RERh$_2$Si$_2$ compounds, the magnetic structure of EuRh$_2$Si$_2$ is composed of ferromagnetic Eu~layers in the ab planes stacking antiferromagnetically along the c~axis. Reflections corresponding to an incommensurate propagation vector (0~0~$\tau$) along the c~axis are clearly seen in resonant magnetic x-ray scattering at the Eu L$_3$~edge below $T_\mathrm{N}$ ~\cite{13}.
Here we report on the strong ferromagnetic properties of the surface and subsurface region in the antiferromagnet EuRh$_2$Si$_2$, which are driven by the ordered local 4f~moments of Eu. These properties are monitored by angle-resolved photoelectron spectroscopy (ARPES) by looking at the diamond-shaped surface state that exists around the $\overline{\mathrm{M}}$-point of the surface Brillouin zone (BZ). This Shockley-type surface~state is observed at the Si-terminated surface, and resides inside a large gap in the projected bulk band structure. The surface ferromagnetism is manifested by a huge splitting of this state due to the exchange~interaction with the ordered local~moments of the Eu~atoms lying three atomic~layers below the surface. The exchange~splitting of the surface states, which may be described as a trapped two-dimensional electron gas within the top four layers, provides immediate information on the magnetism of the first buried Eu~layer in EuRh$_2$Si$_2$ when followed, for instance, as a function of temperature. 

\subsection{ARPES insight into the surface ferromagnetism}
\begin{figure}
  \centering
  \includegraphics[width=0.45\textwidth]{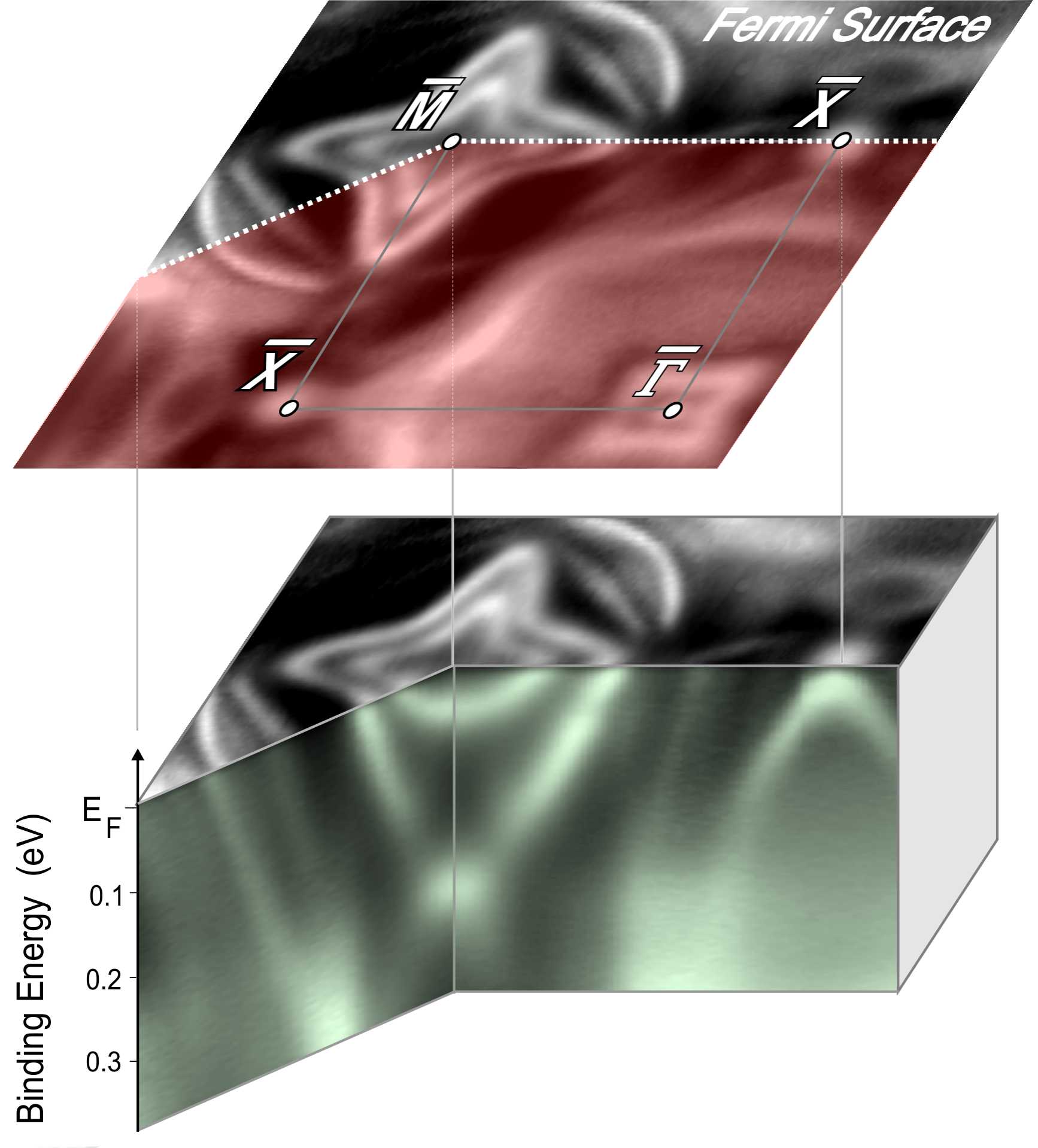}
  \caption{\textbf{Fermi surface and electron bands for the AFM phase of EuRh$_2$Si$_2$.} Three-dimensional representation of the ARPES-derived electron band structure measured at 11~K for a Si-terminated surface of EuRh$_2$Si$_2$ around the $\overline{\mathrm{M}}$-point of the surface BZ.}
  \label{fig:1}
\end{figure}
In Fig.\ref{fig:1} we show the Fermi surface and the electron band structure for the Si-terminated surface of EuRh$_2$Si$_2$ as seen in ARPES at low temperatures ($T = 11$~K). The most explicit feature here is a diamond-shaped surface state, lying inside a large~gap of the projected bulk band~structure around the $\overline{\mathrm{M}}$~point of the surface~BZ. Like in YbRh$_2$Si$_2$~\cite{9}, this Shockley-like surface state is created mainly by Si~3s,~3p (60\%) and Rh~4d (40\%) states, and is intrinsic to the silicon-terminated (001)~surface. However, unlike in YbRh$_2$Si$_2$, the surface state in EuRh$_2$Si$_2$ reveals a clear and strong~splitting. Because of this difference a Rashba~type spin-orbit interaction seems not to be the origin of the observed splitting, as it should be present in YbRh$_2$Si$_2$ as well. A more plausible scenario of the detected splitting is a magnetic exchange~interaction of the surface state with Eu~4f moments in the first buried Eu~layer below the surface. Note that an exchange type splitting of the surface state implies ordered local~moments in the first Eu~layer which is in line with the observed in-plane (ab) ferromagnetic order~\cite{13}. A related phenomenon has been reported for the helical antiferromagnet holmium~\cite{14} with the main difference that there the Ho d-states couple to the 4f~moments on the same site, while in the present case the surface state of the outermost Si-Rh-Si plane couples to Eu-moments in the 4$^{\mathrm{th}}$~subsurface layer. 
\begin{figure}
  \centering
  \includegraphics[width=0.45\textwidth]{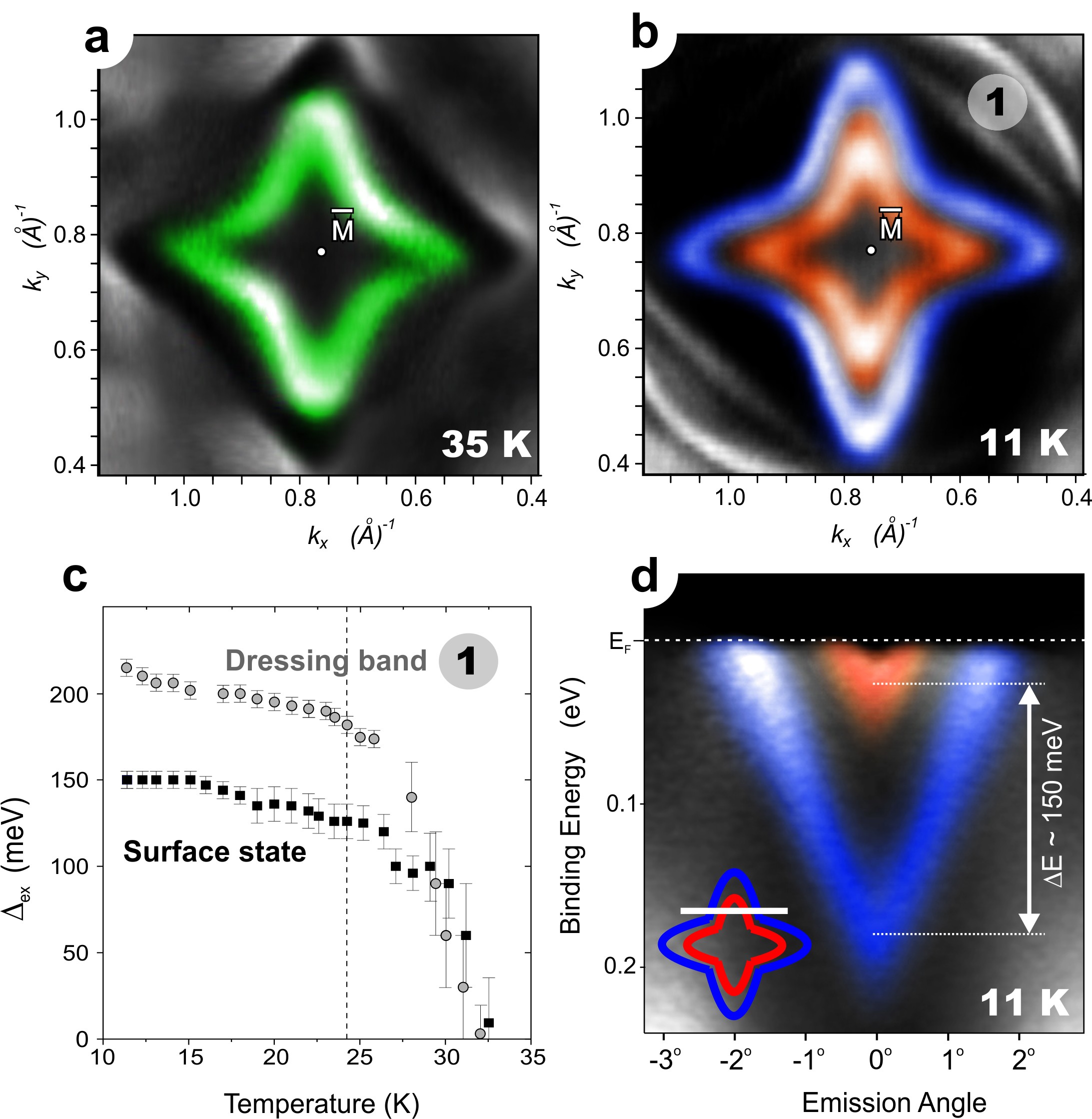}
  \caption{\textbf{ARPES insight into the magnetically induced splitting of the diamond-shaped surface state.} Fermi surface maps taken for the Si-terminated surface of EuRh$_2$Si$_2$ near the $\overline{\mathrm{M}}$-point at 35~K (a) and at 11~K (b), i. e. above and below the bulk AFM transition at $T_\mathrm{N} = 24.5$~K, respectively. (c)~The~ARPES-derived temperature evolution of the magnitude of the surface state (dark symbols) and dressing band’s (light symbols) splitting. For both, the splitting vanishes above approx. 32.5~K. (d)~The~ARPES-derived band map taken at 11~K, demonstrating the largest splitting of the surface state. The inset schematically shows the direction of the measurements.}
  \label{fig:2}
\end{figure}
The assignment of the surface state splitting to exchange~interaction with a magnetic sublayer is supported by measurements performed on a freshly cleaved crystal at $T = 35$~K, i.e. above the Néel temperature. Indeed, no splitting of the surface state is observed at this temperature, as seen in Figs.~\ref{fig:2}a and \ref{fig:2}b, where we show the experimentally observed Fermi surfaces around the $\overline{\mathrm{M}}$~point at high and low temperatures, respectively. When we follow the evolution of the surface state with decreasing temperature (Fig.~\ref{fig:2}c), we can see that the splitting sets in rather sharply at around $32.5$~K and rapidly reaches a value of about $150$~meV (Fig.~\ref{fig:2}d) where it levels off. Note that the onset of the splitting is notably above the bulk ordering temperature, $T_{\mathrm{N}} = 24.5$~K, indicated by the dashed line. The temperature evolution of the splitting was determined by looking at a cut parallel to, but slightly away from the high symmetry direction $\overline{\mathrm{X}}$--$\overline{\mathrm{M}}$ (Fig.~\ref{fig:2}d). This prevents that the second, hole-like surface state with its maximum at about 0.3~eV binding energy overlaps with the features of interest. The temperature dependence for the splitting of this ``dressing'' band is presented in Fig.~\ref{fig:2}c too, and will be discussed below.

\subsection{Origin of the Shockley surface state splitting}
If we ascribe the magnetic splitting of the surface state to exchange interactions with the localized 4f~moments of the topmost Eu~layer, then the size of the splitting represents a direct measure of the magnetization of the Eu~plane in the crystal. The fact that the splitting disappears only notably above the bulk Néel temperature suggests that a certain in-plane order of the Eu~4f moments in the topmost Eu~layer persists even above the bulk ordering temperature. Two scenarios could explain this phenomenon: magnetic fluctuations or a static ferromagnetic order in the topmost Eu~layer.
Magnetic fluctuations above the bulk ordering temperature are a well-known phenomenon and occur on a timescale that is usually large compared to the characteristic timescale of the photoemission process of 10$^{-16}$~s~\cite{15, 16}. With certain probability magnetically ordered domains could then be observed in the photoemission experiment which may explain the experimental results and reflect predominantly a bulk magnetic property. However, the temperature dependence of the splitting, which is proportional to the magnetization, is rather reminiscent of the power-law behavior observed close to a second order phase transition (see Fig.~\ref{fig:2}c). Therefore, the magnetic fluctuation scenario seems unlikely, and there are some indications for a specific magnetic phase close to the surface.
The persistence of static magnetic order in the topmost Eu~layer even above the bulk ordering temperature is a more likely scenario. Such an effect is known, for instance, from ferromagnetic Gd metal where ferromagnetic order in the outermost surface~layer is observed up to $60-80$~K above the bulk Curie temperature~\cite{17,18,19} and reveals all features of an extraordinary phase transition~\cite{20}. In Gd this phenomenon seems to be related to the existence of a weakly dispersing surface state of d$_{\mathrm{z}^2}$-symmetry which reveals a characteristic Stoner-like splitting that scales with the magnetization of the outermost Gd~plane~\cite{19}. The Gd system thus seems to be similar to EuRh$_2$Si$_2$, with the difference that in the latter the responsible magnetic moments are not located directly at the surface but in the 4th subsurface layer and the surface state is not formed by RE derived orbitals, but mainly by Si 3s, 3p and Rh 4d states of the topmost Si-Rh-Si layers. 
 
\subsection{Theoretical insight into the surface and bulk related magnetism}
\begin{figure}[!h]
  \centering
  \includegraphics[width=0.45\textwidth]{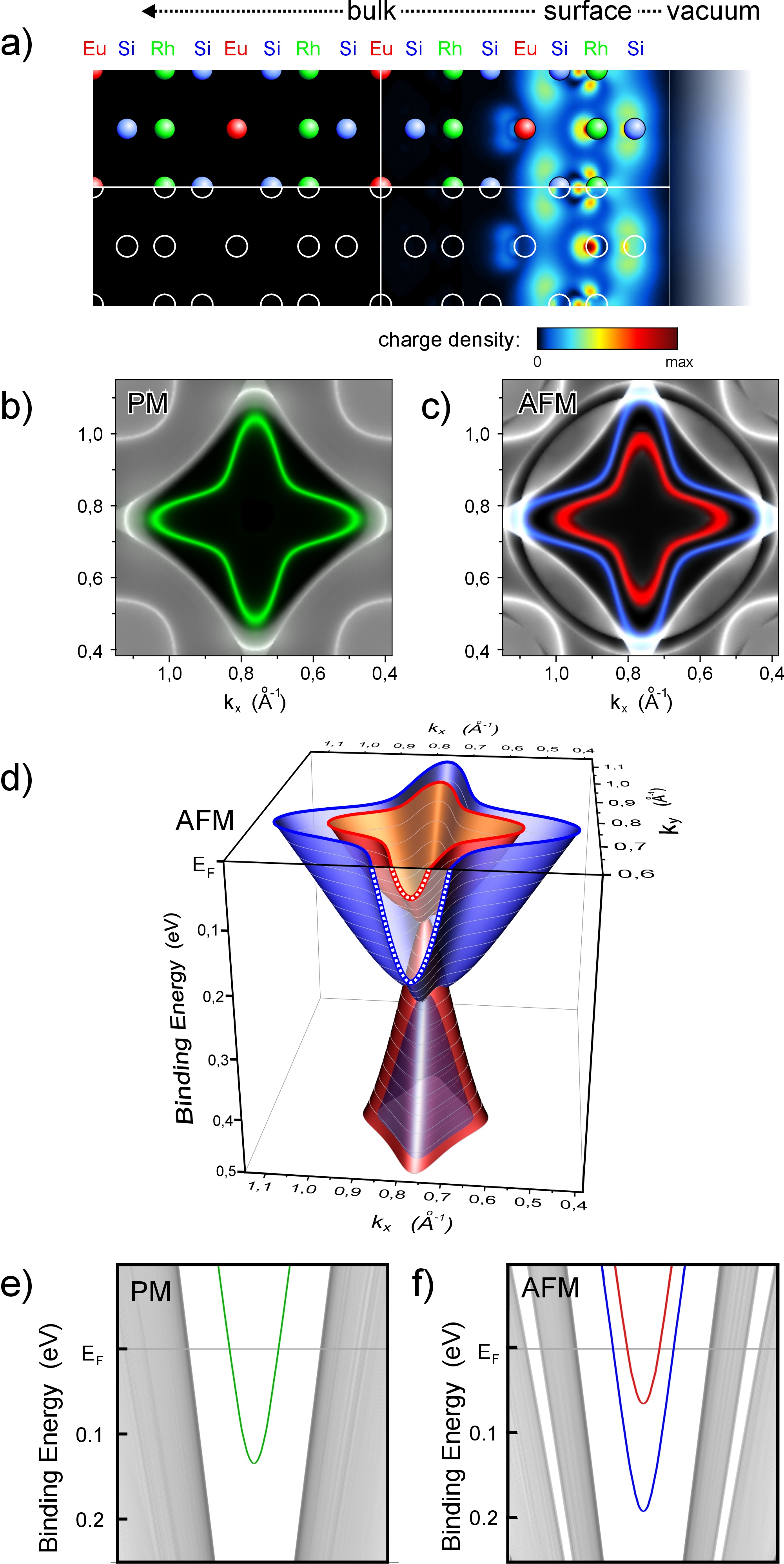}
  \caption{\textbf{Theoretical insight into surface and bulk related electrons in EuRh$_2$Si$_2$.}  (a)~Averaged probability density distribution of the Kohn-Sham eigenstates (projected on the ac~plane) corresponding to the surface~states at the $\overline{\mathrm{M}}$-point obtained for the PM~phase, which is superimposed with the slab used for band-structure calculations. (b) and (c) The computed and superimposed Fermi~surfaces for bulk and slab performed for the PM and AFM phases, respectively. (d) Three-dimensional presentation of the surface state calculated for the AFM phase. (e) and (f) The computed electron band structures along the line (inset in Fig.~\ref{fig:2}d) used in the experimental study of the temperature evolution of the surface state spin split.}
  \label{fig:3}
\end{figure}
To investigate the possibility of such an exchange interaction across four atomic layers we performed a theoretical analysis of the discussed experimental findings. To this end, we calculated and compared the band structures in the surface region assuming several magnetic configurations including the bulk PM and AFM phases. 
In Fig.~\ref{fig:3} we present the results of respective calculations for the latter two configurations. In particular, Fig.~\ref{fig:3}a shows the spatial extension of Kohn-Sham eigenstate corresponding to the surface state at the $\overline{\mathrm{M}}$-point that has been found in the slab calculations. Clearly, the surface state is not strictly confined to the first atomic layer but rather extends over the first four layers up to the Eu ions. Figs.~\ref{fig:3}b and \ref{fig:3}c show the calculated bulk Fermi surface superimposed with that calculated for the slab in the PM and AFM phases, respectively, in close agreement with the ARPES spectra depicted in Fig.~\ref{fig:2}a and \ref{fig:2}b. The bulk related calculations revealed for both phases a similar gap around the $\overline{\mathrm{M}}$-point, while the slab results nicely reproduce the discussed surface state which lies inside this gap and clearly splits in the AFM phase. Note that a similar splitting is also obtained for the case that only the topmost Eu layer is ordered ferromagnetically while the bulk is already in the paramagnetic phase. This provides strong evidence that the observed splitting in the ARPES experiment is indeed caused by the magnetic exchange interaction, lifting up the spin degeneracy of the surface state. Figure~\ref{fig:3}d shows a three-dimensional plot of both electron- and hole-like surface states lying in the gap around the $\overline{\mathrm{M}}$-point. The exchange splitting shows a strong anisotropy with a difference of about 25~meV between the $\overline{\mathrm{X}}$--$\overline{\mathrm{M}}$ and the $\overline{\Gamma}$--$\overline{\mathrm{M}}$ directions.   Figs.~\ref{fig:3}e and \ref{fig:3}f present a similar cut, taken through the surface state parallel to $\overline{\mathrm{X}}$--$\overline{\mathrm{M}}$ as in Fig.~\ref{fig:2}d, for the PM and AFM phases, respectively. Even quantitatively, the exchange splitting obtained from the calculations agrees well with our experimental findings ($\approx 125$~meV vs. $\approx 150$~meV).

\subsection{``Dressing'' band and origin of its splitting}
Interestingly, a closer look at the experimental and theoretical results reveals that a splitting in the AFM phase is not only observed for the prominent surface state but also for the sharp ”dressing” feature that surrounds it and is labeled ”1” in Fig.~\ref{fig:2}b. This band is reproduced in the bulk band structure calculations and should, therefore, reflect bulk properties of the material. The band splitting reveals a similar temperature dependence to that of the surface state (Fig.~\ref{fig:2}a), which implies that it is also governed by magnetism. However, in contrast to the surface state, this splitting is linked to the symmetry breaking caused by the antiferromagnetic order. This is e.g. demonstrated by a calculation in which a superstructure is produced by replacing the divalent Eu atoms of every second Eu-layer by trivalent Gd ions. The results reveal a similar gap formation to that of pure EuRh$_2$Si$_2$ in its AFM phase, although both the Eu and Gd ions were treated in the paramagnetic state. The fact that the gap persists, like the splitting of the surface state, well above the Néel temperature reflects then obviously a phenomenon that is not restricted to the topmost Eu layer alone but extends at least over several Eu layers and is perhaps characteristic for the bulk. The temperature dependence of the dressing band’s splitting (see Fig.~\ref{fig:2}c) favors an interpretation in terms of a second order phase transition over an order parameter fluctuation scenario, indicating that the properties of the near-surface region differ from those of the bulk.
In summary, we have given clear evidence for a large exchange coupling of a Si-derived Shockley surface state to the outermost Eu layer that is located four atomic layers below the surface of the antiferromagnetic compound EuRh$_2$Si$_2$. The resulting exchange splitting provides direct information on the temperature dependent magnetism in the discussed Eu layer and reveals an ordering temperature of 4f moments close to the surface that is notably higher than the bulk TN. Our results suggest that the mechanism of formation of the surface ferromagnetism discovered in EuRh$_2$Si$_2$ can be extended to other antiferromagnetic metallic or semiconducting compounds where surface states exist in an energy gap at the Fermi level. These split surface states may induce magnetization in functional surface layers of topological insulators or/and Rashba type surface systems deposited onto the antiferromagnetic material, thus lifting spin degeneracy of the topological and Rashba type surface states.

\subsection{Acknowledgments}
This work was supported by the DFG (grant VY64/1-1, GE602/2-1, GRK1621). SP acknowledges financial assistance from the Alexander von Humboldt Foundation, Germany.

\subsection{Methods}
\textbf{Experiment.} Angle-resolved photoelectron spectroscopy (ARPES) studies were carried out at the Swiss Light Source (SIS X09LA instrument) as described in detail in Ref.~\citep{12}. The spectra were acquired using a Scienta R4000 electron energy analyzer. The overall energy and angular resolutions were 10~meV and 0.1, respectively. High quality single-crystalline samples of EuRh$_2$Si$_2$ were cleaved in situ in ultra-high vacuum at a base pressure better than $8\times 10^{-11}$~bar. Surface regions terminated by a Si layer were selected by minimizing the Eu~4f surface-related PE signal as the beam was scanned across the sample. For the Fermi surface measurements a new sample was cleaved at each temperature. The temperature dependent measurements were performed always going from high to low temperatures in order to avoid fast sample aging. 

\textbf{Theory.} The electron band structure of EuRh$_2$Si$_2$ was calculated using density functional theory within the local density approximation \cite{21} in a full-potential local orbital basis \cite{22}. If not explicitly mentioned, the scalar-relativistic approximation has been employed. The 4f~basis states of Eu have been fixed to the experimentally found occupation ($n_{\mathrm{4f}} \approx 7.0$) using either an un-polarized, paramagnetic (PM) configuration or a local moment scenario of 7~$\upmu$B per Eu~atom and a ferromagnetic coupling within the Eu~ab planes and AFM interaction along the c~axis. The Si-terminated surface of the EuRh$_2$Si$_2$ crystal has been simulated by a slab that breaks the translational invariance along the c~axis and has at least 7~successive Eu~layers in the conventional unit cell. A detailed description can be found in the supplementary information of \cite{12}. The k-mesh was set to ($12\times 12\times 12$) for the bulk and ($12\times 12\times 1$) for the slab calculations, respectively.

\end{document}